\documentclass{elsart}

\usepackage{amsfonts,amssymb}

\usepackage[dvips]{epsfig}

\newcommand{\green}[1]{\left\langle\!\left\langle #1
                          \right\rangle\!\right\rangle}
\newcommand{\expect}[1]{\left\langle #1 \right\rangle}
\newcommand{\comm}[1]{\left[ #1 \right]}
                         
\begin{document}
\begin{frontmatter}
\title{Thickness dependent Curie temperatures of ferromagnetic Heisenberg films}
\author{R.~Schiller and W.~Nolting}
\address{Humboldt-Universit\"at zu Berlin, Institut f\"ur Physik,
            Invalidenstra{\ss}e 110,\\ D-10115 Berlin, Germany}
\begin{abstract} 
We develop a procedure for calculating the magnetic properties of a
ferromagnetic Heisenberg film with single-ion anisotropy which is valid for
arbitrary spin and film thickness. Applied to sc(100) and fcc(100) films
with spin $S=\frac{7}{2}$ the theory yields the layer dependent 
magnetizations and Curie temperatures of
films of various thicknesses making it possible to investigate magnetic
properties of films at the interesting 2D-3D transition.
\end{abstract}
\end{frontmatter}

In the past the Heisenberg model in thin films and superlattices 
has been subject to intense theoretical work. 
Haubenreisser et al.~\cite{HBCC72} obtained good results
for the Curie temperatures of thin films \cite{DLN79} introducing
an anisotropic exchange interaction (\ref{aei}). Shi and Yang \cite{SY92}
calculated the layer-dependent magnetizations of ultra-thin $n$-layer 
films with single-ion anisotropy (\ref{sia}) for thicknesses $n\leq6$. 
Other recent works are aimed at the question of reorientation transitions in
ferromagnetic films \cite{MJB94} or low-dimensional quantum Heisenberg
ferromagnets \cite{Yab91}.

When investigating the temperature dependent magnetic and electronic properties 
of thin local-moment films or at surfaces of real substances it becomes desirable to be 
able to calculate the magnetic properties of the underlying Heisenberg model
with no restrictions to neither the film thickness $n$ nor the spin $S$ of the
localized moments. We develop a straigthforward analytical approach for the case
of Heisenberg film with single-ion anisotropy. 

Considering the Heisenberg model,
\begin{equation}
\label{ham_or}
\mathcal{H}_f=\sum_{ij} J_{ij} \; \mathbf{S}_i \cdot \mathbf{S}_j
=\sum_{ij} J_{ij} \left( S^+_i S^-_j + S^z_i S^z_j \right) ,
\end{equation}
in a system with film geometry one comes to the conclusion that due to the
Mermin-Wagner theorem \cite{MW66} the problem cannot have a solution 
showing collective magnetic order at finite temperatures $T>0$.

To steer clear of this obstacle there are two possibilities.
First, one can apply a decoupling scheme to the Hamiltonian (\ref{ham_or}) which
breaks the Mermin-Wagner theorem. The most common example in the case of the
Heisenberg model would be a mean-field decoupling.
For us, the main drawback of the mean-field 
decoupling is its incapability of describing physical properties at the 2D-3D
transition.

When choosing a better decoupling approximation to fulfill the Mermin-Wagner
theorem, the original Heisenberg Hamiltonian (\ref{ham_or}) has to be extended
to break the directional symmetry. The most common extensions are the
introduction of an anisotropic exchange interaction,
\begin{equation}
\label{aei}
- D \sum_{ij} S^z_i S^z_j - D_{\mathrm{s}} \sum_{i,j \in \mathrm{surf}} S^z_i S^z_j ,
\end{equation}
and/or the single-ion anisotropy,
\begin{equation}
\label{sia}
- D_0 \sum_i \left( S^z_i \right)^2 - D_{0,\mathrm{s}} \sum_{i \in \mathrm{surf}} 
\left( S^z_i \right)^2 .
\end{equation}
In (\ref{aei}) and (\ref{sia}) the first sums run over all lattice sites of the
film whereas in the second optional terms the summations include positions within the
surface layers of the film, only, according to a possible variation of the
anisotropy in the vicinity of the surface.

Extending the original Heisenberg Hamiltonian (\ref{ham_or}) by 
(\ref{aei}) or (\ref{sia}) one can now calculate the magnetic
properties of films at finite temperatures within a nontrivial decoupling
scheme. 

For the following we have choosen a single-ion anisotropy which is uniform
within the whole film leaving us with the total Hamiltonian:
\begin{equation}
\label{ham_com}
\mathcal{H}=\mathcal{H}_f + \mathcal{H}_A
=\sum_{ij\alpha\beta} J_{ij}^{\alpha\beta} 
\left( S^+_{i\alpha} S^-_{j\beta} + S^z_{i\alpha} S^z_{j\beta} \right) + 
D_0 \sum_{i\alpha} \left( S^z_{i\alpha} \right)^2 ,
\end{equation}
where we have considered the case of a film built up by $n$ layers
parallel to two infinitely extended surfaces.
Here, as in the following, greek letters $\alpha$, $\beta$, ..., indicate the
layers of the film, while latin letters $i$, $j$, ..., number the sites within a
given layer. Each layer possesses two-dimensional translational symmetry. Hence,
the thermodynamic average of any site dependent operator $A_{i\alpha}$ depends
only on the layer index $\alpha$:
\begin{equation}
\expect{A_{i\alpha}} \equiv \expect{A_{\alpha}} .
\end{equation}

To derive the layer-dependent magnetizations $\expect{S^z_{\alpha}}$ for
arbitrary values of the spin $S$ of the localized moments we introduce the 
so-called retarded Callen Green function \cite{Cal63}:
\begin{equation}
\label{callengreen}
G^{\alpha\beta}_{ij(a)} (E) \equiv \green{S^+_{i\alpha} ; B^{(a)}_{j\beta}}_E
=\green{S^+_{i\alpha} ; \mathrm{e}^{a S^z_{j\beta}} S^-_{j\beta}}_E .
\end{equation}
For the equation of motion of the Callen Green function,
\begin{equation}
\label{eom0}
E \; G^{\alpha\beta}_{ij(a)}(E)=\hslash
\expect{\comm{S^+_{i\alpha} , B^{(a)}_{j\beta}}_-} +
\green{\comm{S^+_{i\alpha}, \mathcal{H}}_- ; B^{(a)}_{j\beta} }_E
\end{equation}
one needs the inhomogenity,
\begin{equation}
\label{inhom}
\expect{\comm{S^+_{i\alpha} , B^{(a)}_{j\beta}}_-}=
\eta^{(a)}_{\alpha} \delta_{\alpha\beta} \delta_{ij} ,
\end{equation}
and the commutators,
\begin{eqnarray}
\label{comm_hf}
\comm{S^+_{i\alpha},\mathcal{H}_f}_- &=& -2\hslash \sum_{k\gamma}
J^{\alpha\gamma}_{ik} \left( S^z_{i\alpha} S^+_{k\gamma} -
S^z_{k\gamma} S^+_{i\alpha} \right) , \\
\label{comm_ha}
\comm{S^+_{i\alpha},\mathcal{H}_A}_- &=& D_0\hslash 
\left( S^+_{i\alpha} S^z_{i\alpha} + S^z_{i\alpha} S^+_{i\alpha} \right) .
\end{eqnarray}

For the higher Green function on the right hand side of the equation of motion
(\ref{eom0}) resulting from the commutator relationship (\ref{comm_hf})
one can apply the Random Phase Approximation (RPA) which has proved to yield
reasonable results throughout the entire temperature range:
\begin{eqnarray}
\label{rpa1}
\green{S^z_{i\alpha} S^+_{k \gamma} ; B^{(a)}_{j\beta}}_E & \longrightarrow &
\expect{S^z_{\alpha}} \green{S^+_{k \gamma} ; B^{(a)}_{j\beta}}_E ,\\
\label{rpa2}
\green{S^z_{k \gamma} S^+_{i\alpha} ; B^{(a)}_{j\beta}}_E & \longrightarrow &
\expect{S^z_{\gamma}} \green{S^+_{i\alpha} ; B^{(a)}_{j\beta}}_E .
\end{eqnarray}
For the higher Green functions resulting from the commutator (\ref{comm_ha})
this is not possible due to the strong on-site correlation of the corresponding
operators. However, one can look for an acceptable decoupling of the form
\begin{equation}
\label{lines}
\green{ S^+_{i\alpha} S^z_{i\alpha} + S^z_{i\alpha} S^+_{i\alpha} ; 
B^{(a)}_{j\beta}}_E=\Phi_{i\alpha} 
\green{S^+_{i\alpha} ; B^{(a)}_{j\beta}}_E. 
\end{equation}
As was shown by Lines \cite{Lin67} 
an appropriate coefficient $\Phi_{i\alpha}=\Phi_{\alpha}$ can be found
for any given function $B^{(a)}_{j\beta}=f \left( S^-_{j\beta} \right)$,
which is all we need to know for the moment. We will come back to the explicit
calculation of the $\Phi_{\alpha}$ later.

Using the relations (\ref{inhom})--(\ref{lines}) and applying a two-dimensional
Fourier transform introducing the in-plane wavevector $\mathbf{k}$ the equation
of motion (\ref{eom0}) becomes
\begin{eqnarray}
\label{eom1}
\left( E -\hslash D_0 \Phi_{\alpha} \right) G^{\alpha\beta}_{\mathbf{k} (a)}
&=&\hslash \eta^{(a)}_{\alpha} \delta_{\alpha\beta}\nonumber \\
& & + 2\hslash
\sum_{\gamma} \left( J^{\alpha \gamma}_{\mathbf{0}} \expect{S^z_{\gamma}} 
G^{\alpha\beta}_{\mathbf{k} (a)} - J^{\alpha \gamma}_{\mathbf{k}} 
\expect{S^z_{\alpha}} G^{\gamma\beta}_{\mathbf{k} (a)} \right) .
\end{eqnarray}
Writing equation (\ref{eom1}) in matrix form one immediately gets the solution
by simple matrix inversion:
\renewcommand{\arraystretch}{0.3}
\setlength{\arraycolsep}{0.5ex}
\begin{equation}
\label{sol}
G^{\alpha\beta}_{\mathbf{k} (a)} (E)=\hslash \left( \begin{array}{ccc} 
\eta^{(a)}_1 & & 0 \\ & \ddots & \\ 0 & & \eta^{(a)}_n \end{array} \right)
\cdot \left( E \mathbb{I} - \mathbb{M} \right)^{-1} ,
\end{equation}
where $\mathbb{I}$ represents the $n \times n$ identity matrix and
\begin{equation}
\label{m}
\frac{\left( \mathbb{M} \right)^{\alpha\beta}}{\hslash}=
\left( D_0 \Phi_{\alpha} + 2 \sum_{\gamma} J_0^{\alpha \gamma}
\expect{S^z_{\gamma}} \right) \delta_{\alpha\beta} - 
2 J_{\mathbf{k}}^{\alpha\beta} \expect{S^z_{\alpha}} .
\end{equation}

The local, i.e.~layer-dependent, 
spectral density, $S^{\alpha}_{\mathbf{k} (a)}=- \frac{1}{\pi} \mathrm{Im}
G^{\alpha\alpha}_{\mathbf{k} (a)}$, can then 
be written as a sum of $\delta$-functions
and with (\ref{sol}) one gets:
\begin{equation}
\label{spectral}
S^{\alpha}_{\mathbf{k} (a)}=\hslash \eta^{(a)}_{\alpha}
\sum_{\gamma} \chi_{\alpha\alpha
\gamma}(\mathbf{k}) \delta \left( E - E_{\gamma}(\mathbf{k}) \right) ,
\end{equation}
where $E_{\gamma}(\mathbf{k})$ are the poles of the Green function (\ref{sol})
and $\chi_{\alpha\alpha \gamma}(\mathbf{k})$ are the weights of these poles in
the diagonal elements of the Green function, 
$G^{\alpha\alpha}_{\mathbf{k} (a)}$. Both, the poles and the weigths can be
calculated e.g.~numerically.

Extending the procedure by Callen \cite{Cal63} from 3D to film 
structures\footnote{The only pre-condition for the extension 
is that the spectral density
has the multipole structure (\ref{spectral})}
one finds an analytical
expression for the layer-dependent magnetizations,
\begin{equation}
\label{sz}
\expect{S^z_{\alpha}}=\hslash \frac{ (1+\varphi_{\alpha})^{2S+1} 
(S-\varphi_{\alpha}) + \varphi_{\alpha}^{2S+1} (S+1+\varphi_{\alpha})}
{\varphi_{\alpha}^{2S+1} - (1+\varphi_{\alpha})^{2S+1}},
\end{equation}
where
\begin{equation}
\label{varphi}
\varphi_{\alpha}=\frac{1}{N_{\mathrm{s}}} \sum_{\mathbf{k}} \sum_{\gamma}
\frac{\chi_{\alpha\alpha \gamma} (\mathbf{k})}{\mathrm{e}^{\beta E_{\gamma} 
(\mathbf{k})} - 1} .
\end{equation}
Here, $N_{\mathrm{s}}$ is the number atoms in a layer and 
$\beta=\frac{1}{k_{\mathrm{B}} T}$. The
poles and weigths in (\ref{varphi}) have to be calculated for the special Green
function $G^{\alpha\alpha}_{\mathbf{k} (a)}$ with 
$a=0$\footnote{The parameter $a$ had been introduced to derive (\ref{varphi})
for arbitrary spin $S$.}. In this case the
Callen Green function (\ref{callengreen}) simply becomes:
\begin{equation}
\label{green}
G^{\alpha\beta}_{ij (0)}=G^{\alpha\beta}_{ij}=
\green{S^+_{i\alpha}; B_{j\beta}^{(0)}} 
=\green{S^+_{i\alpha}; S^-_{j\beta}},
\end{equation}
and, according to (\ref{inhom}),
\begin{equation}
\label{eta}
\eta^{(0)}_{\alpha}=\eta_{\alpha}=2\hslash \expect{S^z_{\alpha}}.
\end{equation}

Having solved the problem formally we are left with explicitly calculating the
coefficients $\Phi_{\alpha}$ of equation (\ref{lines}). Applying the spectral
theorem to (\ref{lines}) for the special case of $a=0$ one gets, using
elementary commutator relations:
\begin{equation}
\label{lines0}
\expect{ S^-_{j\beta} S^+_{i\alpha} (2 S^z_{i\alpha} +\hslash)}
=\Phi_{i\alpha} \expect{S^-_{j\beta} S^+_{i\alpha} }_E. 
\end{equation}

We now define the Green function
\begin{equation}
\label{d}
D^{\beta\alpha}_{ji}=\green{S^-_{j\beta};C_{i\alpha}}_E,
\end{equation}
where $C_{i\alpha}$ is a function of the lattice site. Writing down the
equation of motion of $D^{\beta\alpha}_{ji}$ for the limit $D_0\rightarrow0$,
\begin{equation}
\label{eomd}
E \; D^{\beta\alpha }_{ji}(E)=\hslash
\expect{\comm{S^-_{j\beta} , C_{i\alpha}}_-} +
\green{\comm{S^-_{j\beta}, \mathcal{H}_f}_- ; C_{i\alpha} }_E,
\end{equation}
and decoupling all the higher Green functions using the RPA one arrives after
transformation into the two-dimensional $\mathbf{k}$-space at:
\begin{equation}
\label{sold}
D_{\mathbf{k}}^{\beta\alpha}=\hslash \left( \begin{array}{ccc} 
\expect{\comm{S^-_1 , C_1}_-} & & 0 \\ & \ddots & \\ 0 & & 
\expect{\comm{S^-_n , C_n}_-} \end{array} \right)
\cdot \left( E \mathbb{I} - \mathbb{A} \right)^{-1} ,
\end{equation}
where $\mathbb{A}$ is a matrix which is independent on the choice of 
$C_{i\alpha}$. Now putting $C_{i\alpha}$ in (\ref{d}) 
in turn equal to $S^+_{i\alpha}$ and 
to $S^+_{i\alpha} (2 S^z_{i\alpha} +\hslash)$ and applying the spectral
theorem to equation (\ref{sold}) one eventually gets the relation:
\begin{equation}
\label{relcomm}
\frac{\expect{S^-_{j\beta} S^+_{i\alpha}}}
{\expect{\comm{S^-_{i\alpha}, S^+_{i\alpha}}_-}}=
\frac{\expect{S^-_{j\beta} S^+_{i\alpha} (2S^z_{i\alpha} +\hslash)}}
{\expect{\comm{S^-_{i\alpha}, S^+_{i\alpha}(2S^z_{i\alpha} +\hslash)}_-}}.
\end{equation}
The coefficients $\Phi_{i\alpha}$ are then with (\ref{lines0}) given by
\begin{equation}
\label{phi0}
\Phi_{i\alpha}=\frac
{\expect{\comm{S^-_{i\alpha}, S^+_{i\alpha}(2S^z_{i\alpha} +\hslash)}_-}}
{\expect{\comm{S^-_{i\alpha}, S^+_{i\alpha}}_-}}=
\frac
{2\expect{(S_{i\alpha}^z)^2} -\hslash^2 S (S+1)}
{\expect{S_{i\alpha}^z}},
\end{equation}
where, along with commutator relations, the identity
\begin{equation}
\label{spmsmp}
S^{\pm}_{i\alpha} S^{\mp}_{i\alpha}=\hslash^2 S (S+1) \pm 
\hslash S^z_{i\alpha} - (S^z_{i\alpha})^2
\end{equation}
has been used. To avoid the unknown expectation value 
$\expect{(S^z_{i\alpha})^2}$ we apply the spectral theorem to the spectral
density (\ref{spectral}) with $a=0$ and get using (\ref{eta}):
\begin{equation}
\label{smsp}
\expect{S^-_{\alpha} S^+_{\alpha}}=2\hslash \expect{S^z_{\alpha}}
\frac{1}{N_{\mathrm{s}}}
\sum_{\mathbf{k}} \sum_{\gamma} \frac{\chi_{\alpha\alpha \gamma}({\mathbf{k}})}
{\mathrm{e}^{\beta E_{\gamma}(\mathbf{k})} - 1}
\stackrel{(\ref{varphi})}{=} 2\hslash \expect{S^z_{\alpha}} \varphi_{\alpha}.
\end{equation}
Hence, with (\ref{spmsmp}) and (\ref{smsp}), we get
\begin{equation}
\expect{(S^z_{\alpha})^2}=\hslash^2 S (S+1) -\hslash 
\expect{S^z_{\alpha}} (1 + 2 \varphi_{\alpha}),
\end{equation}
and the coeffcients $\Phi_{\alpha}$ can be written in the convienient form
\begin{equation}
\label{phi}
\Phi_{\alpha}=\frac{ 2\hslash^2 S (S+1) - 3\hslash \expect{S^z_{\alpha}}
(1 + 2 \varphi_{\alpha})}{\expect{S^z_{\alpha}}}.
\end{equation}

Together with (\ref{phi}), the equations (\ref{sol}), (\ref{m}), (\ref{sz}),
and (\ref{varphi}) represent a closed system of equations, which can be solved
numerically.

All the following calculations have been performed for spin $S=\frac{7}{2}$,
applicable to a wide range of interesting rare-earth compounds, and
for an exchange interaction in tight-binding approximation $J=0.01 \mathrm{eV}$
which is uniform within the whole film. 
The case where the exchange integrals in the vicinity of the surfaces are
modified has been dealt with by a couple of authors \cite{WMC}.
The single-ion ansitropy which plays the mere role
of keeping the magnetizations at finite temperatures was choosen $D_0/J=0.01$.

\begin{figure}[htbp]
\epsfig{file=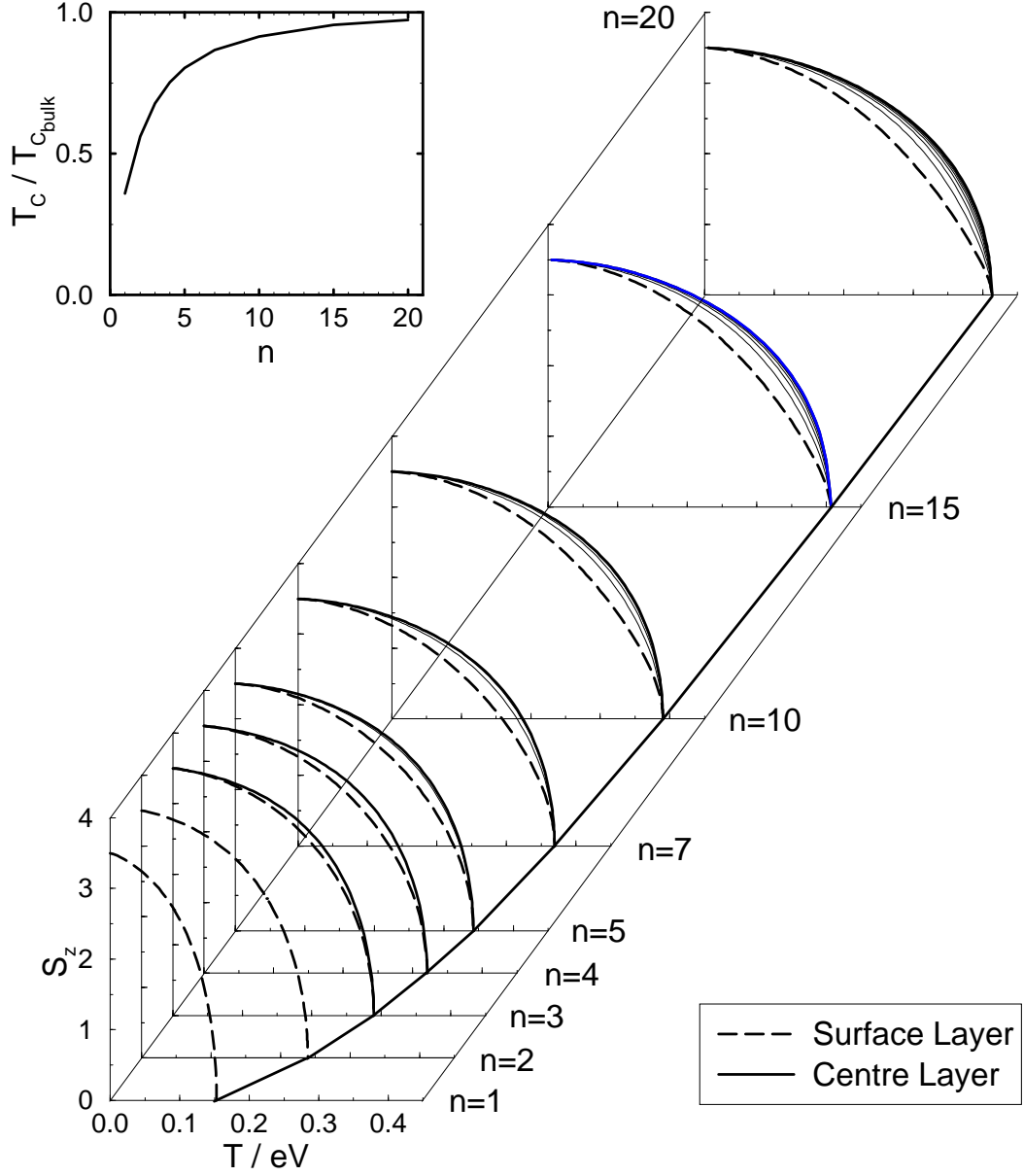, width=\linewidth}
\caption{\label{sc100} Layer-dependent magnetizations, $\expect{S^z_{\alpha}}$,
of sc(100) films as a
function of temperature for various thicknesses $n$. For all temperatures and film
thicknesses the 
$\expect{S^z_{\alpha}}$ increase from the surface layer towards the centre of
the films. {\bfseries Inset:} Curie temperature as a function of
thickness of the sc(100) films.}
\end{figure}

\begin{figure}[htbp]
\epsfig{file=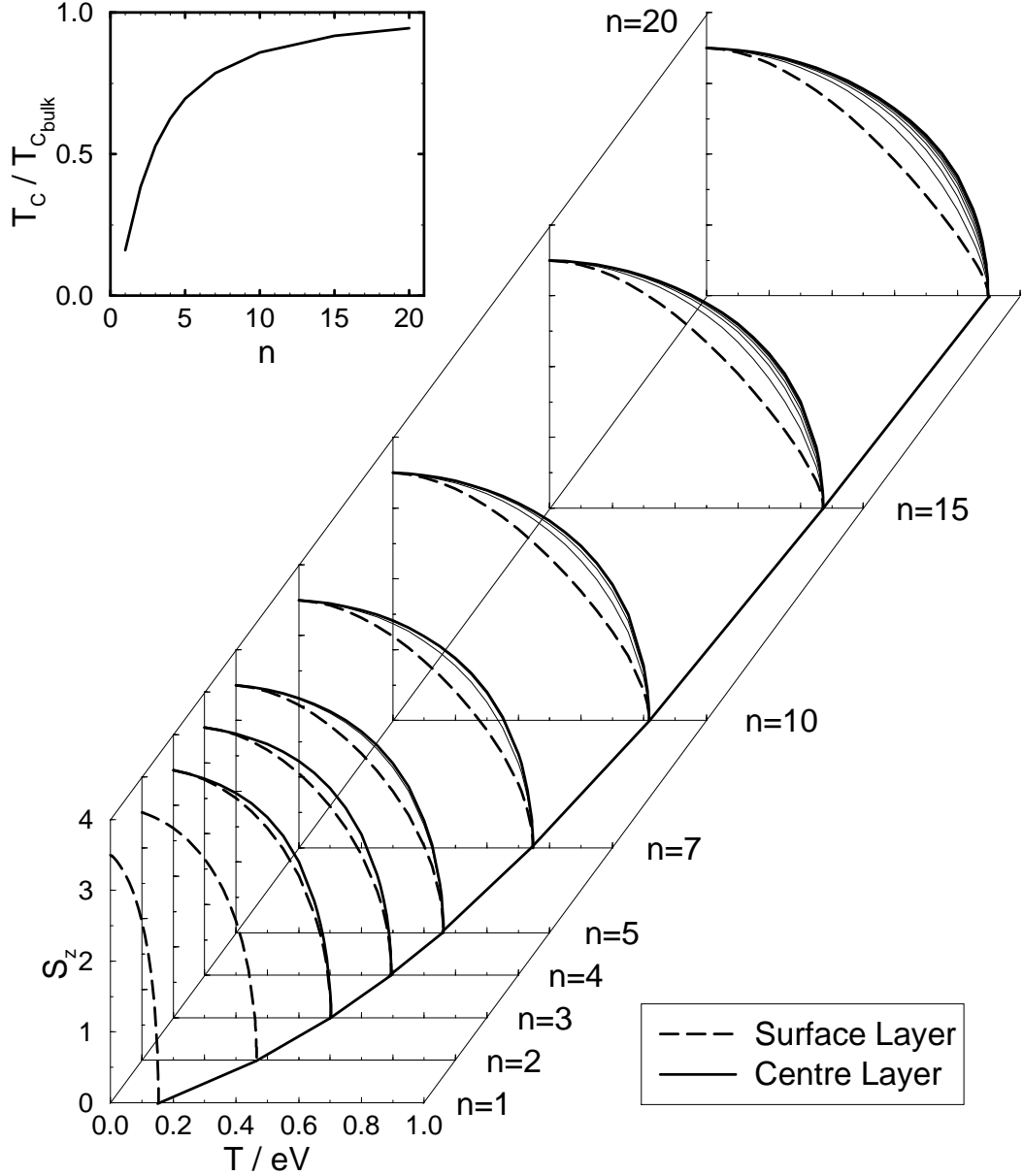, width=\linewidth}
\caption{\label{fcc100} Same as Fig.~\ref{sc100} for fcc(100) films.}
\end{figure}
Figs.~\ref{sc100} and \ref{fcc100} show the temperature and layer-dependent
magnetizations of, respectively, simple cubic (sc) and face-centered cubic (fcc) 
films with the surfaces parallel to the (100)-planes. For the following $Z_s$
means the coordination number of the atoms in the surface layers and $Z_b$ is the
coordination number in the centre layers of the films.
For the case of a monolayer, $n=1$, the curves for the sc(100) and the fcc(100)
'film' are identical, both having the same structure. With increasing film
thickness the Curie temperatures of the films increase. 
For fcc(100) films 
the increase in $T_{\mathrm{C}}$ is steeper resulting in the limit of thick
films in a Curie temperature about twice the value of that of 
the according sc(100) films due to the higher coordination 
number of the fcc 3D-crystal ($Z_{b,fcc}=12$) compared to the sc 3D-crystal 
($Z_{b,sc}=6$).
The larger difference between surface and centre layer
magnetization of the fcc(100) films compared to the sc(100) films can be
explained by the lower ratio between $Z_s$ and $Z_b$
($Z_{s,fcc(100)}/Z_{b,fcc}=8/12$ and $Z_{s,sc(100)}/Z_{b,sc}=5/6$).

Concluding, we have have shown that the presented approach is a useful
and straigthforward method for calculating the layer-dependent magnetizations of
films of various thicknesses and with arbitrary spin $S$ of the localized moments.

\begin{ack}
We would like to thank P.~J.~Jensen for helpful discussions and for bringing
Ref.~\cite{Lin67} to our attention.
One of the authors (R.~S.) would like to acknowledge the support 
by the German National Merit Foundation. The support by the
Sonderforschungsbereich 290 (''Metallische d\"unne Filme: Struktur, Magnetismus
und elektronische Eigenschaften``) is gratefully acknowledged.
\end{ack}

\end{document}